\documentclass[conference]{IEEEtran}
\usepackage[utf8]{inputenc}
\usepackage{mathtools}
\usepackage{url}
\usepackage{xcolor}
\usepackage{array}
\usepackage{bbm}
\usepackage{multirow}
\usepackage{xspace}
\usepackage{makecell}
\usepackage{amsfonts}
\usepackage{bm}
\usepackage[caption = false]{subfig}

\newcolumntype{L}[1]{>{\raggedright\let\newline\\\arraybackslash\hspace{0pt}}m{#1}}
\newcolumntype{C}[1]{>{\centering\let\newline\\\arraybackslash\hspace{0pt}}m{#1}}
\newcolumntype{R}[1]{>{\raggedleft\let\newline\\\arraybackslash\hspace{0pt}}m{#1}}

\DeclarePairedDelimiter{\norm}{\lVert}{\rVert}
\newcommand{\V}[1]{\mathbf{#1}}
\newcommand{\bfx}{\V{x}}
\newcommand{\xad}{\tilde{\bfx}}

\title{(Extended Abstract) Rogue Signs: Deceiving Traffic Sign Recognition with Malicious Ads and Logos}
\author{\IEEEauthorblockN{Chawin Sitawarin, Arjun Nitin Bhagoji, \\ Arsalan Mosenia, Prateek Mittal}
\IEEEauthorblockA{Department of Electrical Engineering\\
Princeton University}
\and
\IEEEauthorblockN{Mung Chiang}
\IEEEauthorblockA{Department of Electrical and Computer Engineering\\
Purdue University}
}

\begin{document}

\newcommand{\signembedding}{Sign Embedding\xspace}
\newcommand{\lentprinting}{Lenticular Printing\xspace}
\newcommand{\adsign}{Logo\xspace}
\newcommand{\customsign}{Custom Sign\xspace}

\maketitle

\begin{abstract}
We propose a new real-world attack against the computer vision based systems of autonomous vehicles (AVs). Our novel \signembedding attack exploits the concept of adversarial examples to modify innocuous signs and advertisements in the environment such that they are classified as the adversary's desired traffic sign with high confidence. Our attack greatly expands the scope of the threat posed to AVs since adversaries are no longer restricted to just modifying existing traffic signs as in previous work. Our attack pipeline generates adversarial samples which are robust to the environmental conditions and noisy image transformations present in the physical world. We ensure this by including a variety of possible image transformations in the optimization problem used to generate adversarial samples. We verify the robustness of the adversarial samples by printing them out and carrying out drive-by tests simulating the conditions under which image capture would occur in a real-world scenario. We experimented with physical attack samples for different distances, lighting conditions and camera angles. In addition, extensive evaluations were carried out in the virtual setting for a variety of image transformations. The adversarial samples generated using our method have adversarial success rates in excess of 95\% in the physical as well as virtual settings.
\end{abstract}



\section{Introduction}\label{sec: intro}
The ubiquity of machine learning (ML) provides adversaries with both opportunities and incentives to develop strategic approaches to fool learning systems and achieve their malicious goals. A number of powerful attacks on the test phase of ML systems used for classification have been developed over the past few years, including attacks on Support Vector Machines and deep neural networks~\cite{biggio2014security,moosavi2015deepfool,szegedy2013intriguing,goodfellow2014explaining,Carlini16}. These attacks work by adding carefully crafted perturbations to benign examples to generate adversarial examples. In the case of image data, these perturbations are typically imperceptible to humans. While these attacks are interesting from a theoretical perspective and expose gaps in our understanding of the working of neural networks, their practical importance remains unclear. The main question to be addressed is ``What is the nature and extent of the attacks that can be carried out on real-world ML systems?'.

Arguably, one of the most important upcoming applications of ML is for autonomous vehicles (AVs) \cite{NVIDIA,bojarski2016end}. Since the computer vision systems of current and future AVs are likely to be based on neural networks \cite{teslaautopilot,applevoxelnet}, the existence of physical world attacks on neural networks would represent a significant threat. However, the attacks on virtual systems listed above do not translate directly to the real world. This occurs because the optimization problems solved to generate virtual adversarial examples do not account for varying physical conditions which may include brightness, orientation and distance variation, camera artifacts, shadows, reflections and the loss of detail from image resizing. Evtimov et al. \cite{Evtimov17} have performed a preliminary investigation of this threat by accounting for some of these factors while creating physical adversarial examples starting from traffic signs. 
\begin{figure}
    \centering
    \subfloat[Benign logo. Classified as 'bicycles crossing' with confidence of 0.59.]{\includegraphics[scale=0.5]{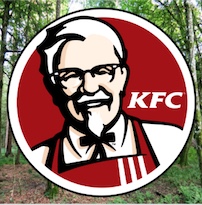}}
    \hspace{1pt}
    \subfloat[Adversarial logo. Classified as 'Stop' with confidence of 1.0. ]{\includegraphics[scale=0.5]{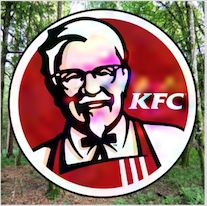}}
    \caption{\textbf{\signembedding attack examples}. The benign image on the left is \emph{rejected} as a false positive detection by our traffic sign recognition pipeline since it has a low confidence. In contrast, the adversarial image on the right is \emph{accepted} since it is classified with high confidence.}
    \label{fig: sign_embed_example}
    \vspace{-13pt}
\end{figure}

\begin{figure*}[t]
    \centering
    \includegraphics[width=\textwidth]{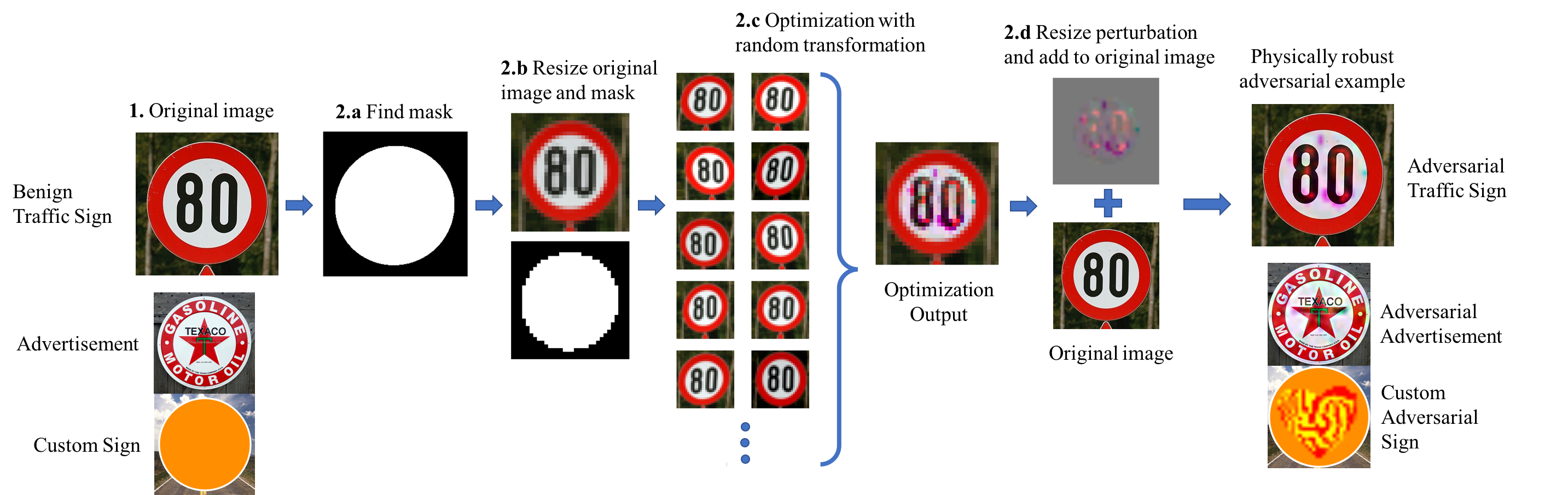}
    \caption{\textbf{Attack pipeline}. This diagram provides an overview of the process by which adversarial examples are generated for the \signembedding attack as well as for adversarial traffic signs.}
    \label{fig: attack}
    \vspace{-10pt}
\end{figure*}

In this paper, we greatly expand the scope of the threat adversarial examples pose to AVs by proposing new attacks to generate physically robust adversarial samples from \emph{innocuous signs} as shown in Figure \ref{fig: sign_embed_example}. We evaluate the real-world viability of these adversarial examples by setting up a realistic evaluation pipeline shown in Figure \ref{fig: attack}. The full version of this paper \cite{sitawarin2018darts} with further details on the methodology and the experiments is available. The code and data required to reproduce our results is available at \url{https://github.com/inspire-group/advml-traffic-sign}. 

The contributions of this paper are as follows: 
\begin{enumerate}
    \item We propose a new real-world attack on traffic sign recognition systems: the \signembedding attack modifies innocuous signs such that they are classified as traffic signs. Our attack pipeline creates adversarial examples which are effective even in a real-world setting.
    \item We propose and examine an end-to-end pipeline for creating adversarial samples that fool sign recognition systems and are resilient to noisy transformations of the image that may occur during the image capture phase. 
    \item We carry out an extensive evaluation of our attacks in both physical as well as virtual settings over various sets of parameters. In the virtual setting, our attack has a 99.07\% success rate without randomized image transformations at test time and 95.50\% with. We also conduct a real-world drive-by test, where we attach a video camera to a car's dashboard and extract frames from the video for classification as we drive by (Figure \ref{fig: real-world_result}). The \signembedding attack has a success rate of over 95\% in this setting, where the success rate is the number of frames in which the adversarial image is classified as the target divided by the total number of frames.
\end{enumerate}

\section{System Model}\label{sec: background}
Machine learning systems typically have two phases, a training phase and a test phase \cite{murphy2012machine}. Our focus is on attacks during the test phase, which are typically known as \emph{evasion attacks}. These have been demonstrated in the virtual setting for a number of classifiers \cite{biggio2013evasion,szegedy2014intriguing,goodfellow2014explaining,Carlini16}. These attacks aim to modify benign examples by adding a perturbation to them such that the modified examples are \emph{adversarial}, i.e. they are misclassified by the ML system. In the case of attacks on the computer vision systems of AVs, the goal of the adversary is to generate signs that appear benign to humans but are misclassified by the traffic sign recognition system.

\noindent \textbf{Threat model}: We consider the commonly used \emph{white-box} threat model \cite{Carlini16,Evtimov17} for the generation of adversarial examples against deep neural networks. In the \emph{white-box} setting, we assume that the adversary has complete access to the traffic sign recognition model including its architecture and weights. Further, we focus on the creation of \emph{targeted} adversarial samples, since these are more relevant to an adversary aiming to misclassify traffic signs.

\noindent \textbf{Virtual adversarial samples}: To generate a \emph{targeted} adversarial sample $\xad$ starting from a benign sample $\bfx$ for a classifier $f$, the following optimization problem \cite{Carlini16} leads to state-of-the-art attack success in the virtual setting:
\begin{align}
\min \quad & d(\xad, \bfx) + \lambda \ell_f(\xad,T),\\
\text{s.t.} \quad & \xad \in C \nonumber.
\end{align}
Here, $\ell_f(\cdot,\cdot)$ is the loss function of the classifier, $d$ is an appropriate distance metric, $T$ is the target class and $C$ is the constraint on the input space. The method described above produces adversarial examples which do not work well under the variety of conditions encountered in the real world. In light of this, there has been some work towards generating physically robust adversarial samples by Athalye et al. \cite{Athalye17} and Evtimov et al. \cite{Evtimov17}. In this paper, we offer a refinement of their methods by incorporating the logit-based objective function and change of variables proposed by Carlini and Wagner \cite{Carlini16} in the virtual setting.


\noindent \textbf{Traffic sign detection and classification}:
Our traffic sign recognition pipeline consists of two stages: detection and classification. We utilize a commonly used recognition pipeline based on the Hough transform \cite{barrile2012automatic, Yakimov2015Hough, Garrido2005Hough}. The shape-based detector uses circle Hough transform \cite{HoughCir83:online} to identify the regions of a video frame that contain a circular traffic sign. Before using Hough transform, we smooth a video frame with a Gaussian filter and then extract only the edges with Canny edge detection \cite{canny}. Triangular signs can be detected by a similar method described in \cite{Yakimov2015Hough}. The detected image patch is cropped and passed on to the neural network classifier to determine whether it is a traffic sign and assign its label. Images classified with a low confidence score are discarded as false positives for detection.


The German Traffic Sign Recognition Benchmark (GTSRB) \cite{Stallkamp2012} is used to train and test the classifier. Our classifier is based on a multi-scale CNN \cite{LeCun2011mltscl} and trained on a data-augmented training set generated by random perspective transformations \cite{Traffics18:online} as well as random brightness and color adjustment of the original training data. The classifier's accuracy on GTSRB validation set is \textbf{98.5\%}. 


\section{Adversarial Examples for Sign Recognition}\label{sec: adv_examples}
In this section, we introduce \signembedding attacks, which modify innocuous signs that are not even part of the training set such that they are detected and classified with high confidence as potentially dangerous traffic signs.

\subsection{\signembedding attacks} \label{ssec:embed}
We propose a novel attack based on the concept of adversarial examples by exploiting the fact that the shape-based detection part of the traffic sign recognition pipeline can pick up a circular object that may not be a traffic sign. Under ordinary conditions when no adversarial samples are present, the false detection does not unduly affect the traffic sign recognition system due to the following reasons; i) The confidence scores corresponding to the predicted labels of these objects are low; ii) These circular objects are not consistently classified as a certain sign. The predicted label changes as the background and the viewing angle varies across multiple frames in the video. Therefore, a traffic sign recognition system can choose to treat any detection with these two properties as an erroneous detection by setting the confidence threshold close to 1 and/or ignoring objects that are inconsistently classified. However, adversarial examples generated from these benign circular objects using our optimization are classified consistently as target traffic signs with high confidence under varying physical conditions. We demonstrate these observations experimentally in Section \ref{subsubsec: validation}.

\begin{figure*}[t]
    \centering
    \includegraphics[width=0.8\textwidth]{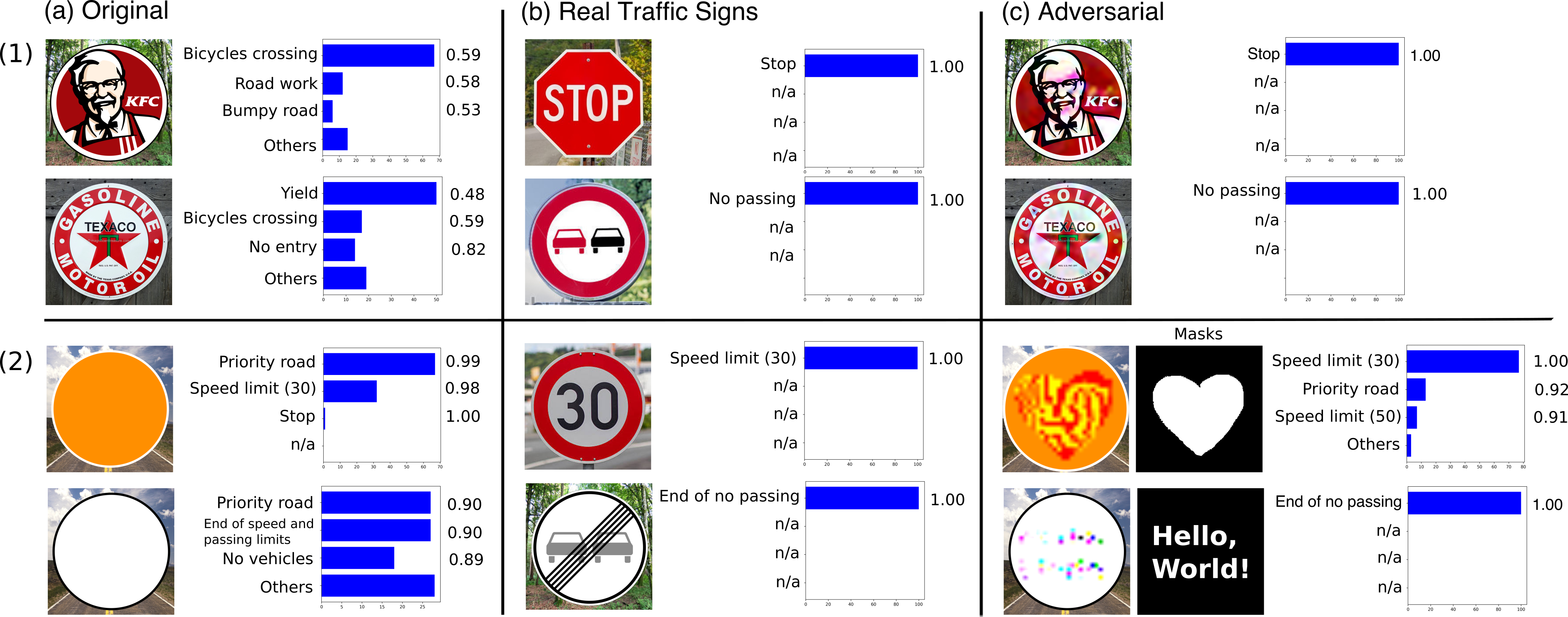}
    \caption{\textbf{Frequency of the top-3 labels the given images are classified as under 100 different randomized transformations}. The numbers on the right of the plot are the average confidence corresponding to those labels. (a) shows original samples. (b) shows real traffic signs of the target class. (c) shows the attacks generated from samples in (a).}
    \label{fig: virtual_adv_samples}
    \vspace{-10pt}
\end{figure*}

\subsubsection{Attack pipeline}
Our pipeline has three steps (Figure \ref{fig: attack}):\\
\noindent \textbf{Step 1.} Obtain the original image $\bfx$ and choose target class $T$ that the adversarial example $\xad$ should be classified as.\\
\noindent \textbf{Step 2.} Generate the digital version of the physically robust adversarial example as follows:

\indent \textbf{1.} Generate mask $M$ for the original image (A mask is needed to ensure that the adversary's perturbation budget is not utilized in adding perturbations to the background.)

\indent \textbf{2.} Resize both the original image and and the mask to the input size of the target classifier ($32\times32$ pixels in our case).

\indent \textbf{3.} Run the optimization from Equation \ref{eq: opt_problem} to obtain the perturbation $\bm{\delta}$.

\indent \textbf{4.} Re-size the output perturbation $\bm{\delta}$ and add it to the original image.

\noindent \textbf{Step 3.} Test and print the generated adversarial signs.

\textbf{Optimization problem}: Our adversarial example generation method involves heuristically solving a non-convex optimization problem using the Adam optimizer \cite{kingma2014adam}. The problem set-up is adapted from a general concept of expectation over transformation \cite{Athalye17}. An updating gradient is averaged over a batch of randomly transformed versions of the original image \cite{Sharif16,Evtimov17}. The robust adversarial example can be written as a solution to the minimization problem given below for any input $\bfx$: 
\begin{align} \label{eq: opt_problem}
\begin{split}
& \min_{\bm{\delta}\in\mathbb{R}^n} \quad c\cdot\frac{1}{B}\sum_{i=1}^{B} \left[ F(\tau_i(\bfx+M\cdot\bm{\delta})) \right] + \max(\norm{\bm{\delta}}_p, L)
\end{split}
\end{align}
where $F(\bfx)=\max(\max_{j \neq T}\{Z(\bfx)_j\} - Z(\bfx)_t, -K)$ is the objective function from Carlini-Wagner's $L_2$ attack \cite{Carlini16} and $Z(x)_j$ is the $j$-th logit (layer before softmax) of the target network. $\tau_i:\mathbb{R}^n\to\mathbb{R}^n$ is a transformation function mapping within the image space ($n=32\times32\times3$). $M$ is a mask or a matrix of zeros and ones with the same width and height as the input image, and $M\cdot\bm{\delta}$ is an element-wise product of the mask and the perturbation to constrain the feasible region to the sign area. The objective value is penalized by a $p$-norm of the perturbation $\bm{\delta}$ in order to keep the adversarial perturbation unnoticeable by humans, and the constant $c$ is adjusted accordingly to balance between the real objective function and the penalty term. The constant $K$ determines the desired objective value and thus, controls \textit{confidence score} of the adversarial example. We introduce an additional constant $L$ to explicitly encourage the norm of the perturbation to be at least $L$ since overly small perturbations can disappear in the process of printing and video capturing.

\noindent \textbf{Image transformations}: For our experiments, the transformation function is comprised of (1) perspective transforms, (2) brightness adjustment and (3) resampling (or image resizing). The three transformations (with randomized parameters) are chosen to simulate the varying real-world conditions in which photos of the adversarial sign might be taken. Perspective transformation covers all projections of a 3D object onto a plane. Thus, it can express common 2D image transformations, such as rotation and shearing.

Using our attack pipeline, the adversary is free to disguise adversarial examples as ad signs, drawings, graffiti etc. Here, we demonstrate two possible settings in which adversarial examples can be embedded: (1) \adsign and (2) \customsign. 

\noindent \textbf{\adsign attack}: In this attack, images of commonly found logos are modified such that they are detected and classified with high confidence as traffic signs (Figure \ref{fig: virtual_adv_samples}(1.a) and (1.c)). Since these logos are omnipresent, they allow an adversary to carry out attacks in a large variety of settings. In this scenario, the problem setup (objective and constraints) is exactly the same as Equation \ref{eq: opt_problem}.

\noindent \textbf{\customsign attack}: In this attack, the adversary creates a custom sign that is adversarial starting from a blank sign (Figure \ref{fig: virtual_adv_samples}(2.a) and (2.c)). This allows the adversary to create adversarial signs in almost any imaginable setting by using a mask to create images or text that are appropriate for the surroundings. In this attack, the original sign is a solid color circular sign and the norm of the perturbation is not penalized by the optimization problem but only constrained by the shape of the mask. This allows the adversary to draw almost any desired shape on a blank sign and the optimization will try to fill out the shape with colors that make the sign classified as the target class. This attack can also be carried out by the same optimization problem by setting $c$ and $L$ to some large numbers so that the optimization will focus on minimizing the losses without penalizing the norm of the perturbation.

\subsubsection{Experimental validation of claims} \label{subsubsec: validation}
To confirm our earlier hypotheses with regard to the confidence of classification for signs out of the training set, we apply random transformations to create 100 images for each of the logo signs. Figure \ref{fig: virtual_adv_samples}(1.a) shows that the logo signs are classified as different classes depending on the transformation and with low confidence on average. As a comparison, Figure \ref{fig: virtual_adv_samples}(b) shows that real traffic signs (one of the $43$ labels) are consistently classified as the correct label with probability very close to 1. 

Some successful \adsign attacks are displayed in Figure \ref{fig: virtual_adv_samples}(1.c). The adversarial signs are classified as their corresponding target label (shown in Figure \ref{fig: virtual_adv_samples}(1.b)) with high confidence. This way, the adversary can expect the desired misclassification with much higher probability and consistency. The embedded adversarial signs are more likely to be recognized by our system as a real traffic sign compared to their original signs because (1) they are classified with a high confidence score and (2) their classified labels are consistent across all frames of the input video. Similarly, in Figure \ref{fig: virtual_adv_samples}(2.c), the \customsign attack produces adversarial signs that contain the shape of the mask filled with various colors. Again, under 100 different randomized transformations, the signs are mostly classified as the target class with high probability.

\noindent \textbf{Remarks}: Our attacks can be created starting from any sign such that they are classified with high confidence as a potentially dangerous traffic sign. Benign signs do not have this effect since they are usually classified with low confidence.

\subsection{Adversarial traffic signs}
Images of traffic signs themselves may be modified to be adversarial, an approach which was followed by Evtimov et al \cite{Evtimov17}. Here, we show that our method can also be used to generate adversarial traffic signs. Our pipeline requires only a single image of a traffic sign, instead of a large number of photos of the target sign taken at different angles and lighting conditions as required by Evtimov et al. \cite{Evtimov17}.

\noindent \textbf{Evaluation in virtual setting}: Using the proposed method, we evaluate our adversarial signs along with those generated by Carlini-Wagner (CW) method on a random subset of 1000 traffic signs chosen from the testing data of GTSRB. Our attack achieves an attack success rate of 99.07\% as compared to 96.38\% for the CW attack. Further, our attack has a much lower deterioration rate of 3.6\% compared to 89.75\% for the CW attack. The deterioration rate is the fraction of adversarial examples that are no longer adversarial after random image transformations are applied to them.



\subsection{Real-world attacks}
\begin{figure}
    \centering
    \subfloat[\adsign attack: classified as a 'No passing' sign with a confidence of 1.0]{\includegraphics[scale=0.24]{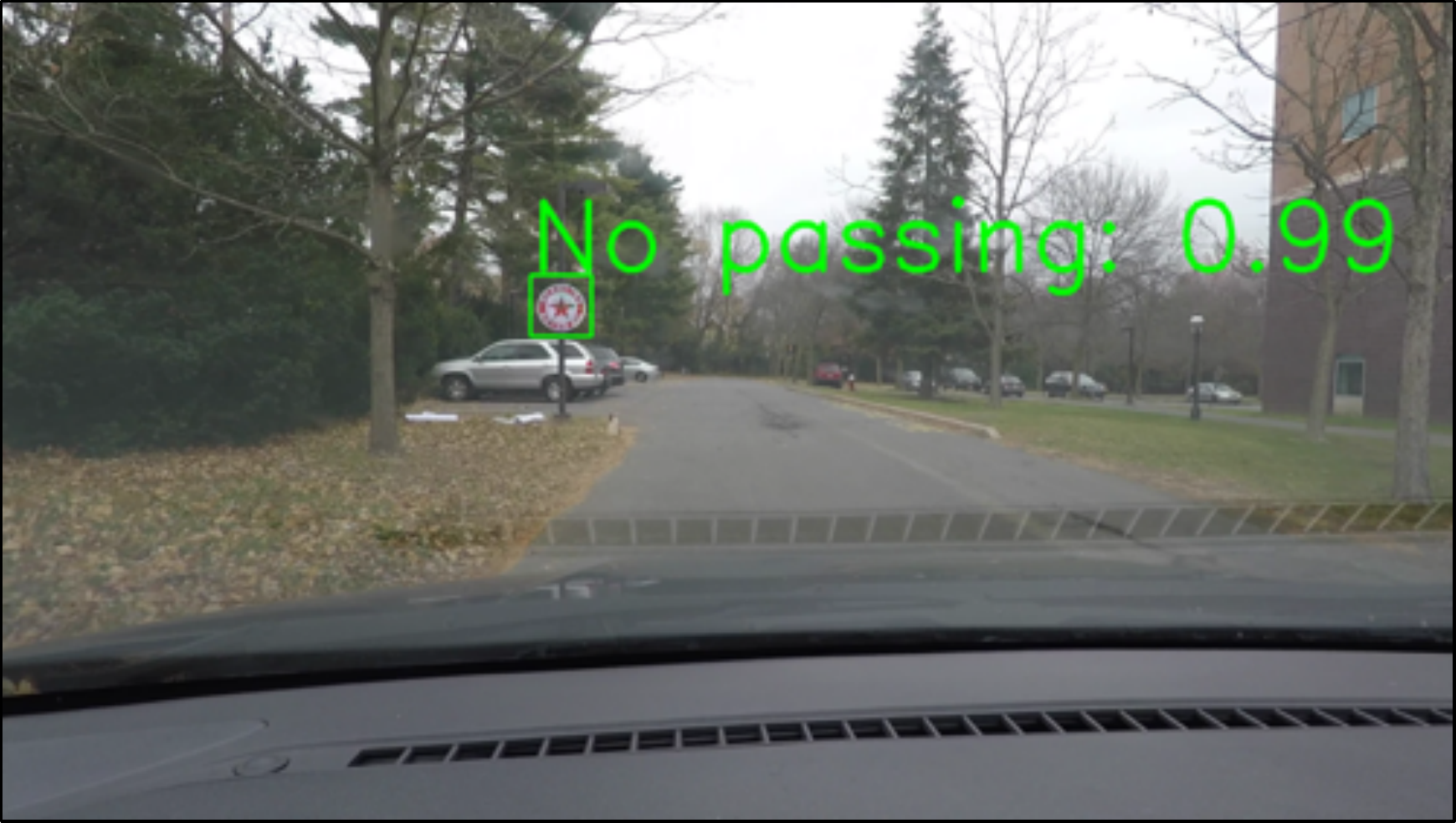}}
    \vspace{0pt}
    \subfloat[\customsign attack: "Hello, World" classified as a 'Stop' sign with a confidence of 1.0]{\includegraphics[scale=0.24]{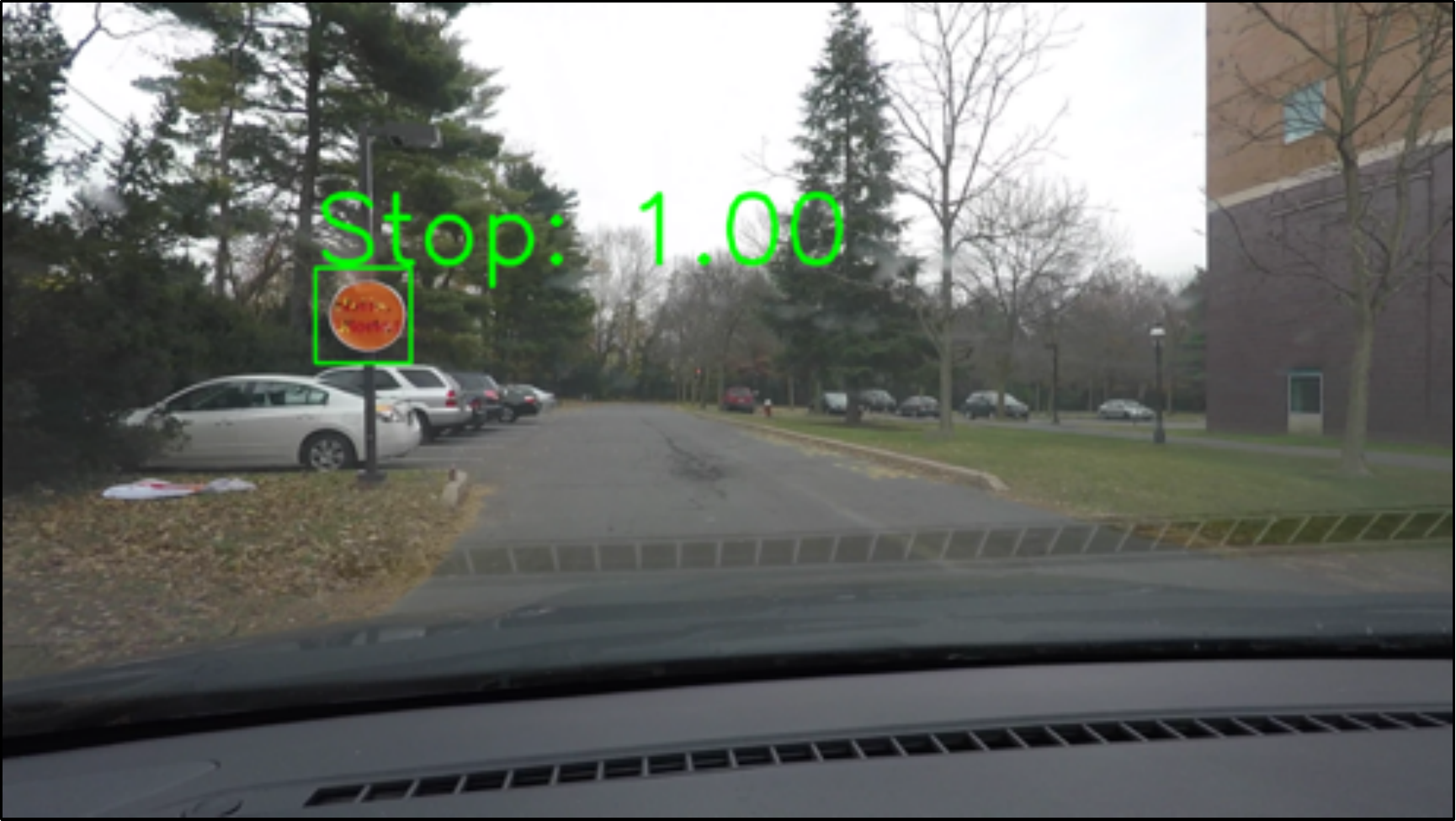}}
    \caption{\textbf{Sample video frames taken during the drive-by tests}}
    \label{fig: real-world_result}
    \vspace{-13pt}
\end{figure}

To demonstrate that the effectiveness of our adversarial traffic signs in the real world, we carried out \emph{drive-by tests} (shown in Figure \ref{fig: real-world_result}) on two samples from each of our attacks (adversarial traffic signs, \adsign, and \customsign). Each sample is re-sized to 30$\times$30 inches and printed on a high-quality poster. The printed signs are stuck on a pole. A GoPro HERO5 was mounted behind the car's windshield to take videos of 2704$\times$1520 pixels at 30 frames per second. Starting around 80 feet from the sign, the car approached it with an approximate speed of 10 mph. One of every five frames is directly passed to the traffic sign recognition pipeline (a combination of a shape-based detector and a CNN classifier). For the adversarial traffic sign, 95.74\% of the detected frames are classified as the adversary's target label, 56.60\% for \adsign attack, and 95.24\% for the \customsign attack.

\section{Conclusion}
We have shown in this paper that the extent of the danger posed by adversarial samples to AVs is more than was previously explored by expanding the attack surface available to an adversary. Finding defenses against our attacks will spur research on creating ML systems that do not provide overly confident predictions on adversarial inputs. We plan to explore these defenses in future work. 

\section*{Acknowledgements}
This  work  was  supported  by  the  National  Science
Foundation under grants CNS-1553437 and CNS-1409415, by Intel through the Intel Faculty Research Award and by the Office of Naval Research through the Young Investigator Program (YIP) Award.



\bibliographystyle{unsrt}
\bibliography{ref.bib}

\end{document}